\documentstyle[aps,preprint,12pt,prb]{revtex}
\tightenlines
\begin{document}
\title{Infrared and Raman spectra of LiV$_2$O$_5$ single crystals}
\author{Z.V.Popovi\'c$^{a}$, R. Gaji\'c$^{b}$, M.J.Konstantinovi\'c$^{c}$,
R. Provoost$^{a}$, V. V. Moshchalkov$^{a}$, A.N.Vasil'ev$^{d}$,
M.Isobe$^{e}$ and Y.Ueda$^{e}$}
\address{ $^a$ Laboratorium voor Vaste-Stoffysica en Magnetism,
Katholieke Universiteit Leuven, Celestijnenlaan 200 D, Leuven
3001, Belgium}
\address{ $^b$ Institute of Physics, 11080
Belgrade, P.O.Box 86, Yugoslavia}
\address{$^c$
Max-Planck-Institut f\"ur
    Festk\"orperforschung,Heisenbergstr. 1,D-70569 Stuttgart, Germany}
\address{ $^d$ Low Temperature Physics Department, Moscow
State University, 119899 Moscow, Russia }
\address{ $^e$
Institute for Solid State Physics, The University of Tokio, 7-22-1
Roppongi, Minato-ku, Tokio 106, Japan}
\maketitle

\begin{abstract}
The phonon dynamics of LiV$_2$O$_5$ single crystals is studied
using infrared and Raman spectroscopy techniques.  The
infrared-active phonon frequencies and dielectric constants are
obtained by oscillator fitting procedure of the reflectivity data
measured at room temperature.  The Raman scattering spectra are
measured at room temperature and at T=10 K in all nonequivalent
polarized configurations.  The assignment of the phonons is done
by comparing the infrared and Raman spectra of LiV$_2$O$_5$ and
NaV$_2$O$_5$. The factor-group-analysis of the LiV$_2$O$_5$
crystal symmetry and of its constituent layers is performed to
explain the symmetry properties of the observed modes.  We
concluded that layer symmetry dominates in the vibrational
properties of this compound.
\end{abstract}
\maketitle

\section{Introduction}

During the past several years low dimensional quantum spin systems
such as the spin-Peierls, spin-ladder and an antiferromagnetic
Heisenberg linear chain systems have attracted much attention
\cite {a1}.  The vanadate family of AV$_2$O$_5$ oxides (A = Li,
Na, Cs, Mg and Ca), which have common V$^{4+}$O$_5$ square
pyramids in the structure, have demonstrated a variety of the low
dimensional quantum spin phenomena:  charge-ordering transition in
NaV$_2$O$_5$ \cite {a2}, spin-gap behavior in Ca(Mg)V$_2$O$_5$
\cite {a3} and CsV$_2$O$_5$ \cite {a4} and typical 1-D behavior
without spin gap in LiV$_2$O$_5$ \cite {a4}. The magnetic
susceptibility data, fitted in the framework of the Bonner-Fisher
model with the exchange interaction J=308 K \cite {a4}, suggest
that the magnetic properties of this system can be described using
homogenous Heisenberg antiferromagnetic linear chain model.  The
nuclear magnetic resonance (NMR) measurements showed the formation
of the staggered spin configurations due to the existence of the
finite-size effect\cite {a5}.

However, in spite of good understanding of the magnetic properties, the study of
the vibrational properties of LiV$_2$O$_5$ is of a great importance because of
the still puzzling interplay between the charge and the magnetic ordering in
NaV$_2$O$_5$ \cite {a6}.  Thus, in this work we present the polarized
far-infrared (FIR) reflectivity as well as the Raman scattering spectra of
LiV$_2$O$_5$ single crystals.  The 16$A_g$, 7$B_{1g}$, 6$B_{3g}$, 9$B_{3u}$, and
5$B_{2u}$ symmetry modes are experimentally observed.  The assignment of the
vibrational modes is done by comparing the phonons in LiV$_2$O$_5$ and
NaV$_2$O$_5$.  Furthermore, our spectra demonstrate the dominance of the layer
symmetry in LiV$_2$O$_5$.

\section{Experiment}

The present work was performed on single crystal plates with
dimensions typically about $1 \times 4 \times 0.5 $ mm$^3$ in the
{\bf a}, {\bf b}, and {\bf c} axes, respectively.  The details of
sample preparation were published elsewhere \cite {a4}.  The
infrared measurements were carried out with a BOMEM DA-8 FIR
spectrometer.  A DTGS pyroelectric detector was used to cover the
wave number region from 100 to 700 cm$^{-1}$; a liquid nitrogen
cooled HgCdTe detector was used from 500 to 1500 cm$^{-1}$.
Spectra were collected with 2 cm$^{-1}$ resolution, with 1000
interferometer scans added for each spectrum.  The Raman spectra
were measured in the backscattering configuration using
micro-Raman system with DILOR triple monochromator including
liquid nitrogen cooled CCD-detector.  An Ar-ion laser was used as
an excitation source.

\section{Results and discussion}

LiV$_2$O$_5$ has an orthorhombic unit cell \cite {a7} with parameters a=0.9702
nm, b=0.3607 nm, c= 1.0664 nm, Z=4 and space group Pnma (D$_{2h}^{16}$).  Each
vanadium atom is surrounded by five oxygen atoms, forming V$O_5$ pyramids.
These pyramids are mutually connected via common edges to form zigzag chains
along the {\bf b}-direction.  Such a crystalline structure is characterized by
two kinds of vanadium double chains along the {\bf b}-axis.  One is magnetic
V$^{4+}$ (S=1/2) and the other one is nonmagnetic V$^{5+}$ (S=0) chain.  These
chains are linked by corner sharing to form layers in (001) - plane, Fig.
\ref{fig1} (a).  The Li atoms are situated between these layers, as shown in
Fig.  \ref{fig1}.1.(b).  The LiV$_2$O$_5$ unit cell consists of four formula
units comprising 32 atoms in all (Fig.  \ref{fig1}).  The site symmetry of all
atoms in Pnma space group is C$_s$.  The factor-group-analysis (FGA) yields
\cite {a8}:

(Li, V$_1$, V$_2$ , O$_1$, O$_2$, O$_3$, O$_4$, O$_5$) (C$_s$):

$\Gamma$ = 2$A_g$ + $A_u$ + $B_{1g}$ + 2$B_{1u}$ + 2$B_{2g}$ + $B_{2u}$ +
$B_{3g}$ + 2$B_{3u}$,

Summarizing these representations and subtracting the acoustic ($B_{1u}$ +
$B_{2u}$ + $B_{3u}$) and silent (8$A_u$) modes, we obtained the following
irreducible representations of LiV$_2$O$_5$ vibrational modes of Pnma space
group:

$\Gamma$ = 16$A_g$(aa,bb,cc) + 8$B_{1g}$(ab) + 16$B_{2g}$(ac) + 8$B_{3g}$(bc) +
15$B_{1u}$({\bf E}$||${\bf c}) + 7$B_{2u}$({\bf E}$||${\bf b}) + 15$B_{3u}$({\bf
E}$||${\bf a})

Thus, 48 Raman and 37 infrared active modes are expected to show up in the
LiV$_2$O$_5$ spectra.  The room temperature polarized far-infrared reflectivity
spectra of LiV$_2$O$_5$ are given in Fig.  \ref{fig2}.  The open circles are the
experimental data and the solid lines represent the spectra computed using a
four-parameter model for the dielectric constant \cite {a9}.

\begin{equation}
\epsilon=\epsilon_{\infty} \prod_{j=1}^{n} \frac{\omega_{LO,j}^2-\omega^2+\imath
\gamma_{LO,j}^{}\omega}{\omega_{TO,j}^2-\omega^2+\imath\gamma_{TO,j}^{}\omega}
\end{equation}

where $\omega_{LO,j}^{}$ and $\omega_{TO,j}^{}$ are longitudinal and transverse
 frequencies
of the j$^{th}$ oscillator, $\gamma_{LO,j}^{}$ and
$\gamma_{TO,j}^{}$ are their corresponding dampings, and
$\epsilon_{\infty}$ is the high-frequency dielectric constant.
The static dielectric constant, given in Table I, is obtained
using the generalized Lyddane-Sachs-Teller relation $\epsilon_0=
\epsilon_{\infty} \prod_{j=1}^{n} \omega_{LO,j}^2 /
\omega_{TO,j}^2$.

The best-oscillator-fit parameters are listed in Table I.  The
agreement between observed and calculated reflectivity spectra is
rather good.  For the {\bf E}$||${\bf a} polarization, nine
oscillators with TO frequencies at about 202, 243, 255, 340, 397,
545, 724, 948, 1006 cm$^{-1}$ are clearly seen.  In the {\bf
E}$||${\bf b} polarization (Fig.  \ref{fig2} (b)) five oscillators
at 180, 255, 323, 557 and 595 cm$^{-1}$ are observed.  We failed
to obtain usefull signal for {\bf E}$||${\bf c} spectra because of
the very small thickness ({\bf c}-axis) of the sample.  The room
and the low temperature Raman spectra of LiV$_2$O$_5$, for
parallel and crossed polarizations, are given in Fig. \ref{fig3}.
The spectra for parallel polarizations consist of $A_g$ symmetry
modes.  Thirteen modes at 100, 123, 172, 209, 328, 374, 398, 528,
550, 639, 725, 965 and 989 cm$^{-1}$ are clearly seen for the (aa)
polarization, two additional modes at 197 and 456 cm$^{-1}$ for
the (bb) polarization and one additional mode at 267 cm$^{-1}$ for
the (cc) polarization.  For the crossed (ab) polarization seven
Raman active $B_{1g}$ symmetry modes at 168, 250, 270, 333, 546,
646 and 737 cm$^{-1}$ are found.  In the case of the (bc)
polarization, six $B_{3g}$ symmetry modes were observed with
almost the same frequencies as $B_{1g}$ modes.  The Raman spectra
for (ac) polarization is given in Fig. \ref{fig3} (f).  For this
polarization ($B_{2g}$ symmetry) we could not resolve any new
mode.  Namely, all modes observed for this polarization have been
already seen in parallel or in other crossed polarizations,
probably because of low-quality of the (010) surface.  The
frequencies of all observed Raman active modes are given in Table
II.

We will first consider the Raman
spectra shown in Fig.  \ref{fig3}.  From the 16 $A_g$ , 8$B_{1g}$, 8$B_{3g}$
and 16$B_{2g}$ modes predicted by FGA of LiV$_2$O$_5$, we clearly observe
16$A_g$, 7$B_{1g}$, and 6$B_{3g}$ modes.  The missing modes seam to be of a very
weak intensity.  The (ac) polarized spectra (Fig.  \ref{fig3} (f)) consist
mostly of $A_g$ modes (and some $B_{1g}$).  The appearance of $B_{1g}$ and
$B_{3g}$ symmetry modes at the same frequencies lead us to consider the crystal
structure of this oxide once again.  The LiV$_2$O$_5$ crystal is a layer crystal
(see Fig.  \ref{fig1} (b)).  The vibrational properties of such crystals
demonstrate the dominance of the layer symmetry \cite {a9,a10}.  The unit cell
of LiV$_2$O$_5$ consists of two layers with 16 atoms in all.  The full symbol of
space group is P 2$_1$/n 2$_1$/m 2$_1$/a.
The first symbol represents $2_1$-screw axis
along {\bf a}-axis followed by diagonal glide plane (n) perpendicular to {\bf a}-
axis with
translation (b+c)/2; the second symbol is the $2_1$ -screw axis parallel to
{\bf b}-axis
with mirror plane perpendicular to it and third symbol means $2_1$-screw axis
parallel to {\bf c}-axis with glide plane perpendicular to {\bf c}-axis with
 translation of
a/2 (see the lower part of Fig.  \ref{fig4}).  If we consider only one layer
we break the periodicity in the direction perpendicular to the layer.  As a
consequence, there are no more symmetry operations along the {\bf a}- and
{\bf c}-axes.  Now
we should consider the layer symmetry in terms of diperiodic groups \cite {a11}
rather than triperiodic space groups.  We found four operation of layer
symmetry, Fig.  \ref{fig4}:

1, identity

$2_1$, twofold screw axis parallel to the {\bf b}-axis

$\overline{1}$, center of symmetry located between VO$_5$ pyramids

m, mirror plane perpendicular to the {\bf b}-axis.

The DG15 diperiodic group is the only symmetry operation group of 80 that has
these symmetry operations.  The full symbol of this diperiodic group is P 1
$2_1$/m 1.  This group is isomorphic with the C$_{2h}^2$ (P$2_1$/m) space group
(second setting) \cite {a12}.  All atoms are in symmetry position (e) of this
space group (with C$_s$ point symmetry).  The normal mode distribution for the
layer is

$\Gamma$=16$A_g$ (xx,yy,zz, xz) +8$B_g$ (xy, yz) +8$A_u$ ({\bf E}$||${\bf y})
+16$B_u$ ({\bf E}$||${\bf x}, {\bf E}$||${\bf z})

The compatibility diagram, relating the layer and the crystal vibration of
LiV$_2$O$_5$, together with the schematic representation of the layer and the
crystal symmetry operations as well as the experimental conditions for the
observation of the optical modes, are given in Fig.  \ref{fig4}.  According
to Fig.  \ref{fig4} the $A_g$ modes for the layer symmetry appear for parallel
and for (ac) crossed polarization and the $B_{1g}$ and $B_{3g}$ modes of crystal
symmetry merge into $B_g$ modes of the layer symmetry.  These facts are fully in
agreement with our Raman spectra, Fig.  \ref{fig3}.  Thus, we can conclude
that the vibrational properties of crystal are predominantly due to vibrational
properties of the layer.  As we already mentioned, the size of samples in (010)
plane and worse quality of this surface do not allow us to check this conclusion
also by infrared spectroscopy.

By comparing the LiV$_2$O$_5$ spectra with the corresponding spectra and the
lattice dynamics of NaV$_2$O$_5$ \cite {a13,a14} we analyze the LiV$_2$O$_5$
phonon properties.  The basic building blocks of LiV$_2$O$_5$ crystal structure
are VO$_5$ pyramids which are mutually connected by edge to build the chains.
Such chains are also present in NaV$_2$O$_5$.  The difference between crystal
structures of Na- and Li- vanadate is illustrated in Fig.  \ref{fig5}.  In
NaV$_2$O$_5$ there is only one vanadium site while in LiV$_2$O$_5$ two
nonequivalent vanadium atom positions are present.  Besides that, the
V$_1$-O-V$_2$ angle in LiV$_2$O$_5$ is smaller (120$^o$) than in NaV$_2$O$_5$
(140$^o$).  Distance between atoms in VO$_5$ pyramids does not differ
significantly in Li- and Na- vanadates.  This produces only a small frequency
shift of the corresponding modes in these compounds.  The existence of the
 two different
VO$_5$ chains with nearly the same inter-atomic distance can
produce the appearance of mode doublets.  Raman spectra for (aa)
and (bb) polarizations of both compounds are shown in Fig.
\ref{fig5}.  As it can be seen from Fig. \ref{fig5}, each mode of
NaV$_2$O$_5$ in the bond stretching region (above 450 cm$^{-1}$)
appears as a doublet in LiV$_2$O$_5$.  The highest intensity mode
in NaV$_2$O$_5$ originates from V-O$_1$ bond stretching vibration.
The distance between these atoms in Na-vanadate is 1.62
$\overcirc{A}$.  In Li vanadate there are two such bonds, one
longer bond 1.65 $\overcirc{A}$ (V$_1$-O$_3$) and one slightly
shorter 1.61 $\overcirc{A}$ (V$_2$-O$_2$) than V-O$_1$ bond in
Na-vanadate.  Thus, we can expect the appearance of two modes with
frequencies higher and lower than the Na V-O$_1$ bond stretching
frequency.  In fact, two modes are observed in Fig.  \ref{fig5} at
lower and higher frequency than Na V-O$_1$ mode.  Applying the
same analysis to the another doublet in bond stretching energy
region we conclude that 528/550 cm$^{-1}$ modes originate from
V$_1$-O$_4$ (V$_2$-O$_5$) bond stretching vibration.  This doublet
corresponds to the mode at 534 cm$^{-1}$ of NaV$_2$O$_5$.

Similar non - phononic broad structure, observed in Na - vanadate
for (aa) polarization at about 640 cm$^{-1}$ (see Fig.
\ref{fig5}), is found in LiV$_2$O$_5$ as a doublet at about
639/725 cm$^{-1}$.
There are several scenarios to explain this structure \cite{a14}.
First, electric dipole transitions between split crystal field
levels, second two-magnon scattering, and third possible
electron-phonon coupled modes observed in antiferromagnets. In the
case of NaV$_2$O$_5$ the wide structure shifts to lower energy
upon cooling \cite{a15}. Fischer {\it et al.}\cite {a14} concluded
that such kind of behavior comes from electron-phonon coupling
like in antiferromagnetic FeCl$_2 $\cite{a16}. The opposite
frequency shift vs. temperature of wide structure maximum in
NaV$_2$O$_5$ then in FeCl$_2$ is connected with the strong spin
fluctuations in low dimensions \cite {a14}.

Two-magnon excitations lead to broad band structures in Raman
scattering spectra. The very recent neutron scattering
measurements \cite{a17} of LiV$_2$O$_5$ show antiferromagnetic
periodicity along the b-axis (chain direction). These results
suggests that LiV$_2$O$_5$ can be regarded as a collection of
independent S=1/2 AF linear chains with zone boundary energy of
$42\pm3$ meV. The frequencies of observed modes (640/725
cm$^{-1}$) are comparable with the values of exchange coupling
energy J=308 K (3J=640 cm$^{-1}$) and zone boundary energy range
(2E$_{ZB}$= 628-724 cm$^{-1}$), respectively. However, the
polarization selection rules allow the appearance of modes for
polarization of incident and scattered light parallel to the
dominant exchange path (b-axis direction). Because we observed
broad modes only for (aa) and (cc) polarization and there is no
frequency shift of these modes to higher energies by lowering of
the temperature we concluded that two-magnon scenario is not
realistic.

Finally, we consider the appearance of these modes as crystal
field induced \cite {a18}. These transitions between different
crystal field levels should be discrete, and no significant
frequency shift vs. temperature can be expected. This is in
accordance with our spectra given in Fig.3. The appearance of two
peaks can be understood due to existence of two structurally
different strongly deformed VO$_5$ pyramids.

The mode at 788 cm$^{-1}$ is probably IR active LO mode (790
cm$^{-1}$, see Table I) but its appearance is not understood at
the moment.

The modes at frequencies below 450 cm$^{-1}$ originate from bond
bending vibrations.  In the Na-vanadate the V-O$_3$-V bending mode
appears at 448 cm$^{-1}$.  The corresponding mode in Li-vanadate
could be the one at 374 cm$^{-1}$.  Note that, the frequency of
this mode is about 16 \% lower than in Na-vanadate, as a
consequence of the change of the bond-bending force constant. We
make an approximate estimation for the change of the bond-bending
force constant by scaling the frequency as the square root of the
force constant, and assuming a scaling of the force constants as
R$^{-6}$ (R is the bond length) \cite {a19}. Similarly, scaling of
the phonon frequency for the bond stretching mode is R$^{-3}$. In
our case the R (O-V-O)$_{Na}$/R(O-V-O)$_{Li}$ is 0.98.  This
parameter can produce the decrease of the phonon frequency of only
5 \%.  The remaining part of frequency decrease probably comes
from modification of bond-bending force constant due to
redistribution of the electron density \cite {a20}.  Namely,
according to Refs.\cite {a21,a22} the electrons in NaV$_2$O$_5$
are located in a V-O$_3$-V molecular orbital.  This is not the
case in LiV$_2$O$_5$ where the electrons are located in V$^{4+}$
chains.

The highest intensity mode for (bb) polarization in Li sample is
the mode at 328 cm$^{-1}$.  This mode originates from O-V-O
bending vibrations as the highest intensity (bb) polarized mode in
Na-sample.  The broad structure at about 398 cm$^{-1}$ seems to
correspond to Li vibration.  If we consider mass effect only, the
replacement of heavier Na atoms with lighter Li produces the shift
of corresponding mode towards higher frequency according to
($\omega_{Na}$=179 cm$^{-1}$) x (m$_{Na}$/m$_{Li}$)$^{1/2}$ = 326
cm$^{-1}$, which is not to far from the observed mode.  Besides,
the Li-O vibrations are expected \cite {a23} at about 390
cm$^{-1}$ .  The pair of modes at 172/209 cm$^{-1}$ originates
from O-V-O bending vibration, and the pair of lowest frequency
modes at 100/123 cm$^{-1}$ represents chain rotation modes.

The similar analysis could be conducted for B$_{1(3)g}$ modes.
Namely, the modes at 646 and 737 cm$^{-1}$ are V$_1$-O$_4$ and
V$_2$-O$_5$ bond stretching vibrations; the three modes at 250,
270 and 333 cm$^{-1}$ are bond bending vibrations, and the lowest
frequency mode at 168 cm$^{-1}$ represents chain rotation mode. We
should mention that similar mode exists in NaV$_2$O$_5$ for
crossed polarization too.

Identification of the infrared modes,
obtained by comparison with IR spectra and lattice dynamics of NaV$_2$O$_5$ is given
in Table I and we do not repeat it here again.

In conclusion, we have measured
the infrared and Raman spectra of LiV$_2$O$_5$.  The assignment of the phonons is done
by comparing the infrared and Raman spectra of LiV$_2$O$_5$ and NaV$_2$O$_5$.  The
factor-group-analysis of the LiV$_2$O$_5$ crystal symmetry and of its constituent
layers is performed to explain the symmetry properties of the observed modes.
Finally, our spectra demonstrate the dominance of the layer symmetry in LiV$_2$O$_5$.

Acknowledgments

Z.V.P.  acknowledges support from the Research Council of the
K.U.  Leuven and DWTC.  The work at the K.U.  Leuven is supported by Belgian
IUAP and Flemish FWO and GOA Programs.  MJK thanks Roman Herzog - AvH for partial
financial support.

\clearpage
\begin{figure}
\caption {Schematic representation of the LiV$_2$O$_5$ crystal structure in
the (a) (001) and (b) (010) plane. }
\label{fig1}
\end{figure}
\begin{figure}
\caption {Room temperature polarized far-infrared reflectivity spectra of
LiV$_2$O$_5$ single crystal for (a) {\bf E}$||${\bf a} and (b) {\bf E}$||${\bf b}
polarizations. The experimental
values are given by the open circles. The solid lines represent the calculated spectra
obtained by fitting procedure described in the text. }
\label{fig2}
\end{figure}
\begin{figure}
\caption {Room and low temperature Raman scattering spectra for all principal
polarized configurations. $\lambda_L$ = 514.5 nm}
\label{fig3}
\end{figure}
\begin{figure}
\caption {Compatibility diagram relating the layer and crystal vibrations in LiV$_2$O$_5$.}
\label{fig4}
\end{figure}
\begin{figure}
\caption {The (aa) (solid lines) and (bb) (doted lines)
polarized room temperature Raman spectra of LiV$_2$O$_5$ and
NaV$_2$O$_5$. Inset: Schematic representation of basing building blocks of Li-
and Na-vanadate  crystal structure.}
\label{fig5}
\end{figure}

\newpage
\narrowtext
\begin{table}
\caption{Oscillator fit parameters in ${\rm cm^{-1}}$.}
\begin{tabular}{cccccccc}
\tableline Polarization & $\omega_{TO}^{}$ & $\gamma_{TO}^{}$ &
$\omega_{LO}^{}$ & $\gamma_{LO}^{}$ & $\varepsilon_0$ &
$\varepsilon_\infty$ & Remark \\ \tableline & 202 & 16 & 230  & 18
& \\ & 243 & 6 & 246 & 6 & {} & {} & ${\rm Li\parallel{\bf a}}$ \\
& 255 & 7 & 268 & 23 & {} & {} & O-V-O bending \\ & 340 & 80 & 380
& 60 & {} & {} & O-V-O bending \\ ${\rm\bf E\parallel a}$ & 397 &
30 & 402 & 80 & 16.7 & 6.5 & O-V-O bending \\ & 545 & 20 & 622 &
15 & {} & {} & ${\rm V_{1(2)}-O_1}$ stretching \\ & 724 & 40 & 727
& 50 & {} & {} & ${\rm V_{1(2)}-O_{4(5)}}$ stretching \\ & 948 & 4
& 962 & 5 & {} & {} & ${\rm V_1-O_3}$ stretching \\ & 1006 & 4 &
1011 & 4 & {} & {} & {} ${\rm V_2-O_2}$ stretching \\ \tableline &
180 & 25 & 207 & 35 & {} & {} & O-V-O bending \\ & 255 & 10 & 259
& 15 & {} & {} & O-V-O bending \\ ${\rm\bf E\parallel b}$ & 323 &
10 & 350 & 15 & 13.4 & 4.5 & ${\rm V_1-O_1-V_2}$ bending \\ & 557
& 23 & 571 & 15 & {} & {} & ${\rm V_{1(2)}-O_1}$ stretching \\ &
595 & 18 & 790 & 20 & {} & {} & ${\rm V_{1(2)}-O_{4(5)}}$
stretching \\ \tableline
\end{tabular}
\end{table}

\newpage
\narrowtext
\begin{table}
\caption{The frequencies (in ${\rm cm^{-1}}$) of Raman active
modes of LiV$_2$O$_5$.}
\begin{tabular}{ccccc}
\tableline Number of peaks & A$_g$ & B$_{1g}$ & B$_{2g}$ &
B$_{3g}$
\\ \tableline 1 & 100 & 168 & 168 & \\ 2 & 123 & 250 & 250 & 123
A$_g$-2 \\ 3 & 172 & 270 & 270 & 172 A$_g$-3 \\ 4 & 197 & 333 &
333 & 250 B$_{1g}$-2
\\ 5 & 209 & 546 & {} & 270 B$_{1g}$-3 \\ 6 & 267 & 646 & 646 &
328 A$_g$-7 \\ 7 & 328 & 737 & 737 & 333 B$_{1g}$-4 \\ 8 & 374 &
{} & {} & 374 A$_g$-8 \\ 9 & 398 & {} & {} & 528 A$_g$-11 \\ 10 &
456 & {} & {} & 550 A$_g$-12 \\ 11 & 528 & {} & {} & 646
B$_{1g}$-6 \\ 12 & 550 & {} & {} & 737 B$_{1g}$-7 \\ 13 & 639 & {}
& {} & 965 A$_g$-15 \\ 14 & 725 & {} & {} & 989 A$_g$-16 \\ 15 &
965 & \\ 16 & 989 & \\  \tableline
\end{tabular}
\end{table}

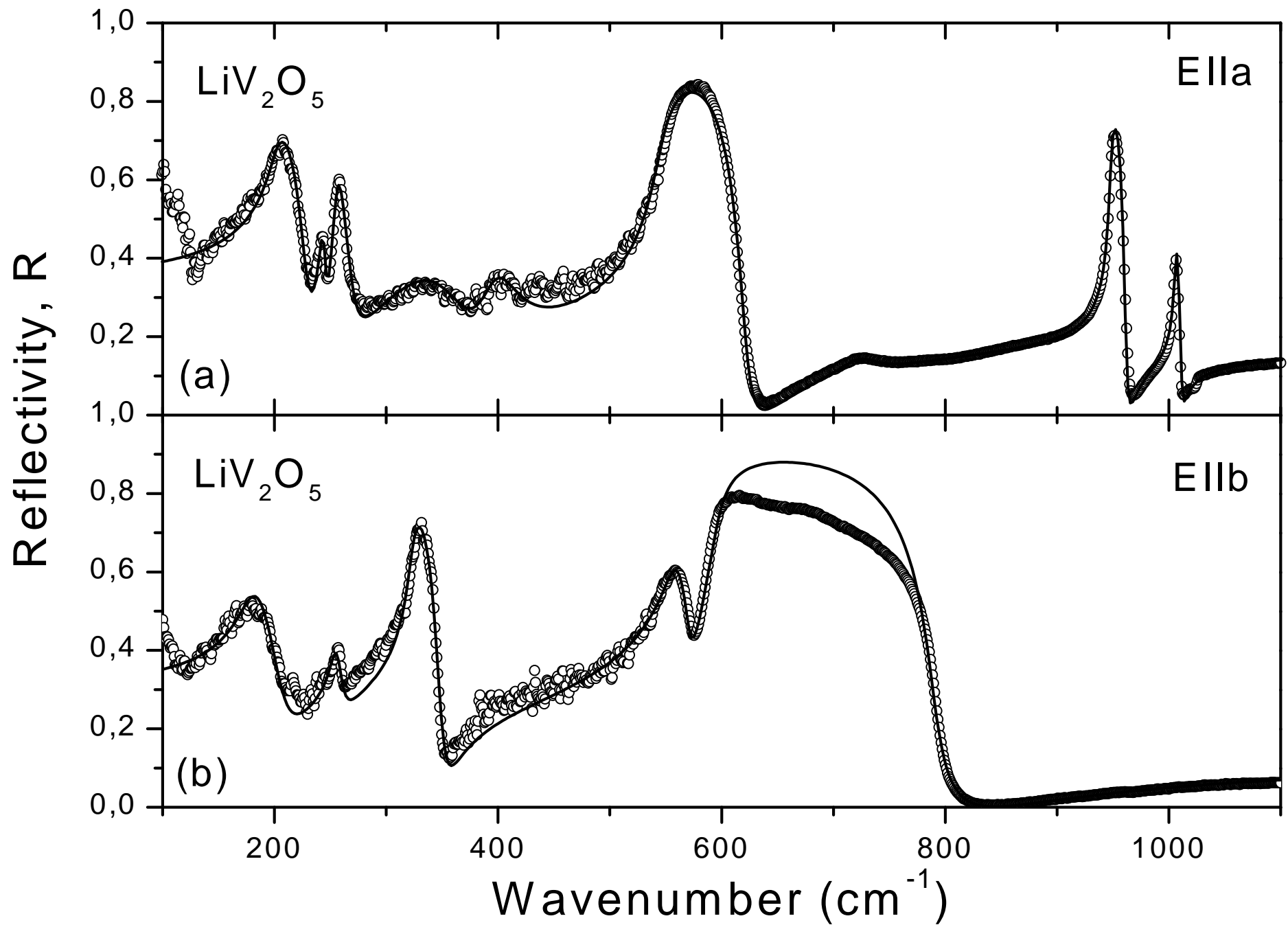
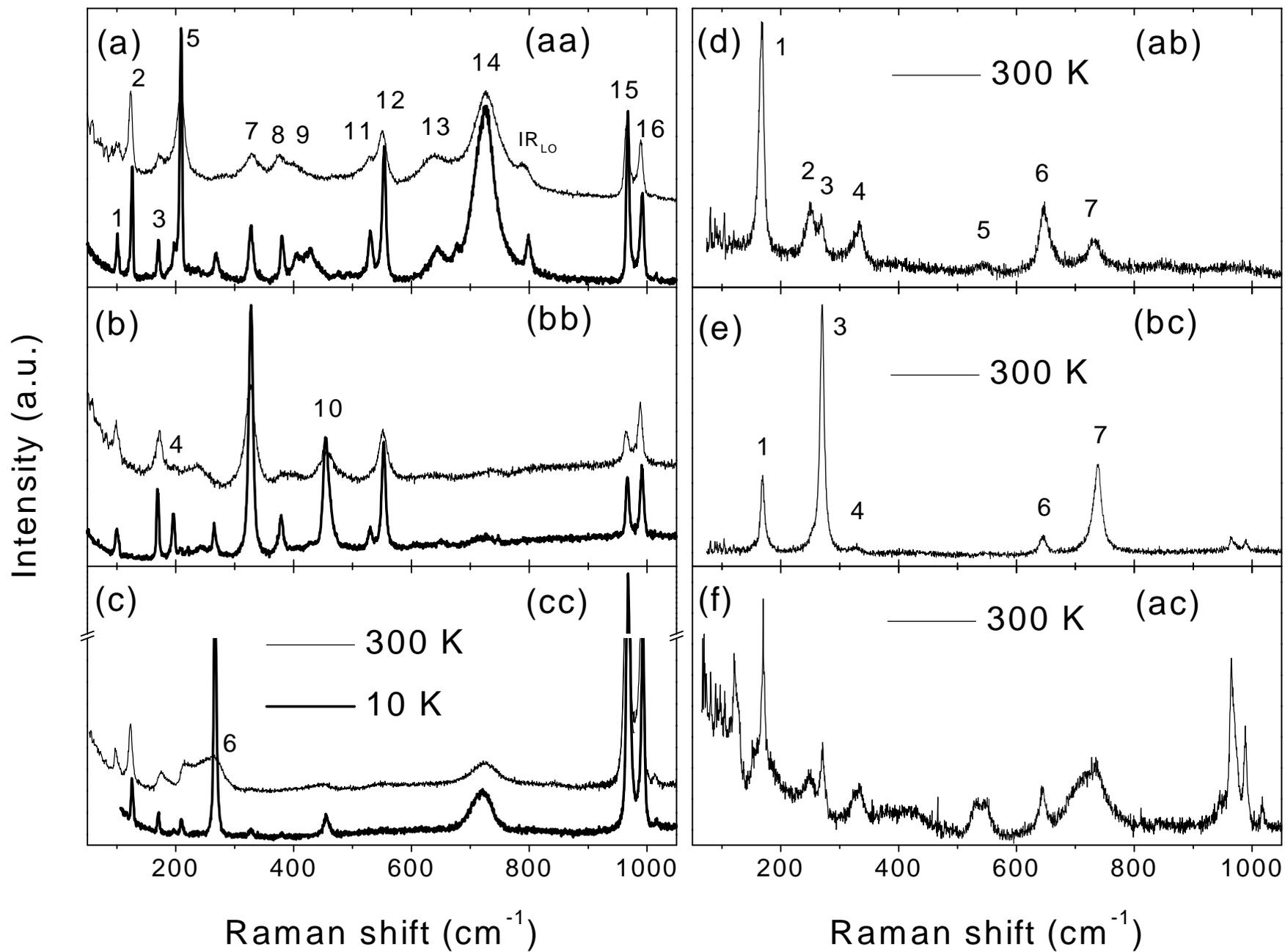
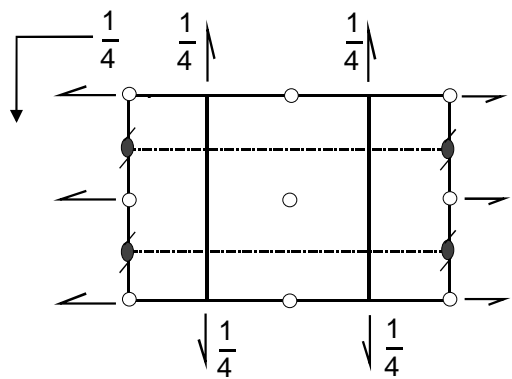
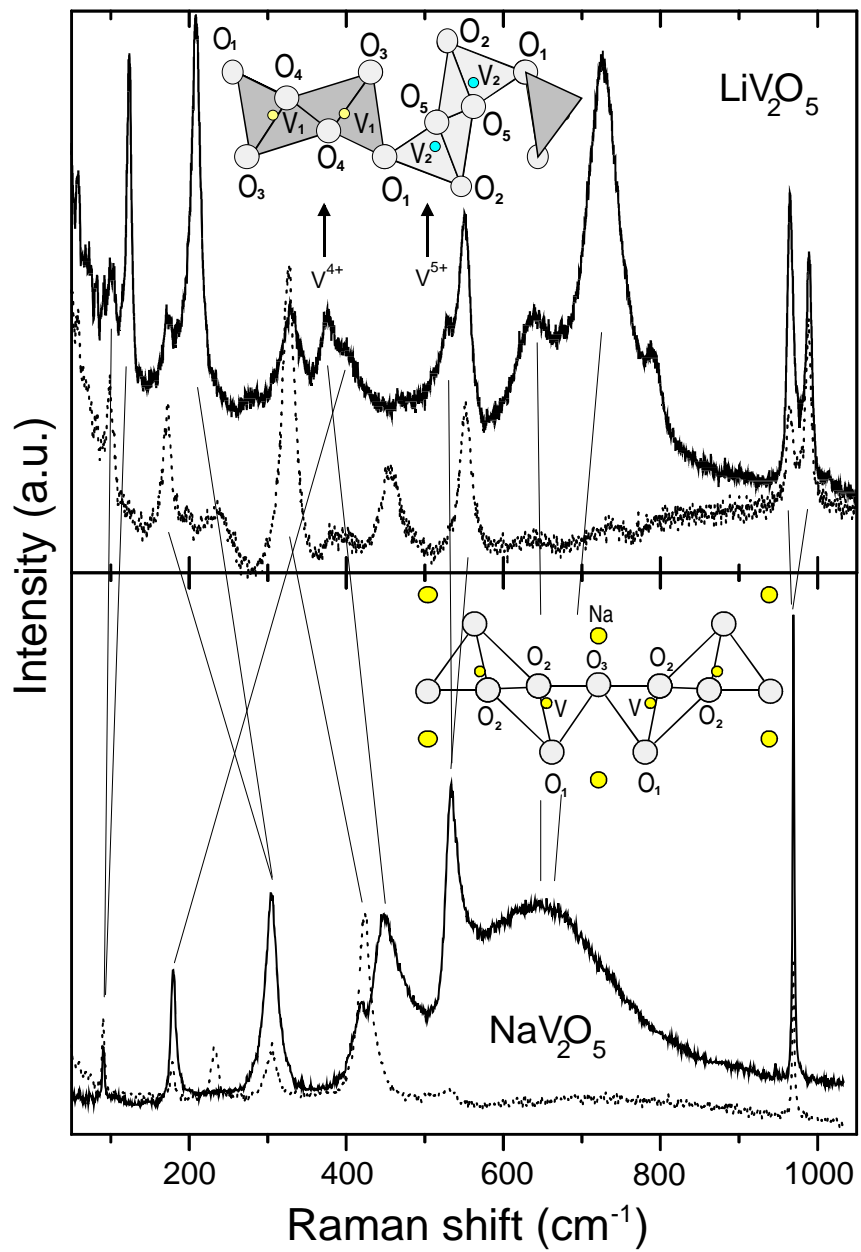
\end{document}